\documentclass[a4paper,10pt,twocolumn,twoside,journal]{article}

\usepackage{geometry}
\advance\oddsidemargin by -1in
\advance\evensidemargin by -1in
\advance\headsep by 2.8mm
\usepackage{listings}
\usepackage{braket}
\usepackage{pgfplots}
\pgfplotsset{compat=1.18}
\usepackage{subcaption}
\usepackage{bm}
\usepackage{placeins}
\usepackage{graphicx} 
\usepackage{color}
\usepackage{setspace}
\usepackage{amsmath}
\usepackage{comment}
\usepackage{algorithm}
\usepackage{algorithmicx}
\usepackage{algpseudocode}
\usepackage{orcidlink}
\usepackage{authblk}

\def\BibTeX{{\rm B\kern-.05em{\sc i\kern-.025em b}\kern-.08em
    T\kern-.1667em\lower.7ex\hbox{E}\kern-.125emX}}

\usepackage[normalem]{ulem}

\title{A Design Framework for the Simulation of\\ Distributed Quantum Computing}

\author{Davide~Ferrari\orcidlink{0000-0002-4777-7234}}
\author{Michele~Amoretti\orcidlink{0000-0002-6046-1904}}
\affil{\small \href{http://www.qis.unipr.it/quantumsoftware.html}{\textit{Quantum Software Laboratory}}, University of Parma\\ \href{mailto:michele.amoretti@unipr.it}{michele.amoretti@unipr.it}, \href{mailto:davide.ferrari1@unipr.it}{davide.ferrari1@unipr.it}}

\date{}

\begin{document}

\maketitle

\begin{abstract}\textbf{
The growing demand for large-scale quantum computers is pushing research on Distributed Quantum Computing (DQC). Recent experimental efforts have demonstrated some of the building blocks for such a design. DQC systems are clusters of quantum processing units (QPUs) connected by means of quantum network infrastructures. Their extension ranges from the single box to the geographical scale. Furthermore, they can be integrated with classical High Performance Computing systems. 
Simulation modeling of DQC architectures provides a safe way to test and explore different what-if scenarios. Many simulation tools have been developed to support the research community in designing and evaluating quantum computer and quantum network technologies, including hardware, protocols, and applications. However, a framework for DQC simulation putting equal emphasis on computational and networking aspects has never been proposed, so far. 
In this paper, a design framework for DQC simulation is presented, whose core component is an Execution Manager that schedules DQC jobs for running on networked quantum computers. Two metrics are proposed for evaluating the impact of the job scheduling algorithms with respect to QPU utilization and quantum network utilization, beyond the traditional concept of makespan. The discussion is supported by a DQC job scheduling example, where two different strategies are compared in terms of the proposed metrics.}
\end{abstract}


\section{Introduction}

The number of qubits that can be embedded in a single quantum chip is limited from the emergence of noise, which is caused for example by changes in the environment, crosstalk, quantum decoherence and implementation errors~\cite{VanDev2016}. 
This hard technological limitation affects all major quantum computing technologies, such as superconducting Josephson junctions, ion traps, quantum dots, etc. As a consequence, both academic and industry communities agree on the need for a quantum computing paradigm-shift toward Distributed Quantum Computing (DQC)~\cite{DQCSurvey2022}, in order to realize large-scale quantum processors. Such a change of approach consists in clustering together multiple quantum processing units (QPUs) by means of a quantum network infrastructure, with the purpose of scaling the number of qubits~\cite{CuoCalCac2020,FerCacAmo2021}.

The realization of the aforementioned vision has already started. For example, IBM is working on a 1386-qubit multi-chip quantum processor, denoted as Kookaburra, to be released in 2025.  
As a demonstration of the quantum communication links supported by this new device, IBM will connect three Kookaburra chips into a 4158-qubit system.

In the near future, one could expect to find quantum processors integrated into High Performance Computing (HPC) facilities and communicating at short distances. In such a scenario, DQC may exploit the already existing low-latency communication technologies (e.g., Omni-Path and InfiniBand) for high-speed classical messaging between quantum processors. Also entanglement distribution, which is highly relevant to DQC, can be achieved over standard telecom fibers connecting the quantum processors~\cite{LagoRivera2021}. To handle DQC jobs submitted by the users, a specialized execution management software would be necessary.

DQC at geographical scale is also expected further into the future, leveraging the Quantum Internet, i.e., metropolitan-area and wide-area quantum networks that work in synergy with the existing Internet~\cite{Wehner2018,HerPomBeu-22}. In this context, long-distance entanglement is enabled by quantum repeaters~\cite{Rakonjac2023}.

\begin{figure*}[ht!]
    \centering
    \includegraphics[width=0.85\linewidth]{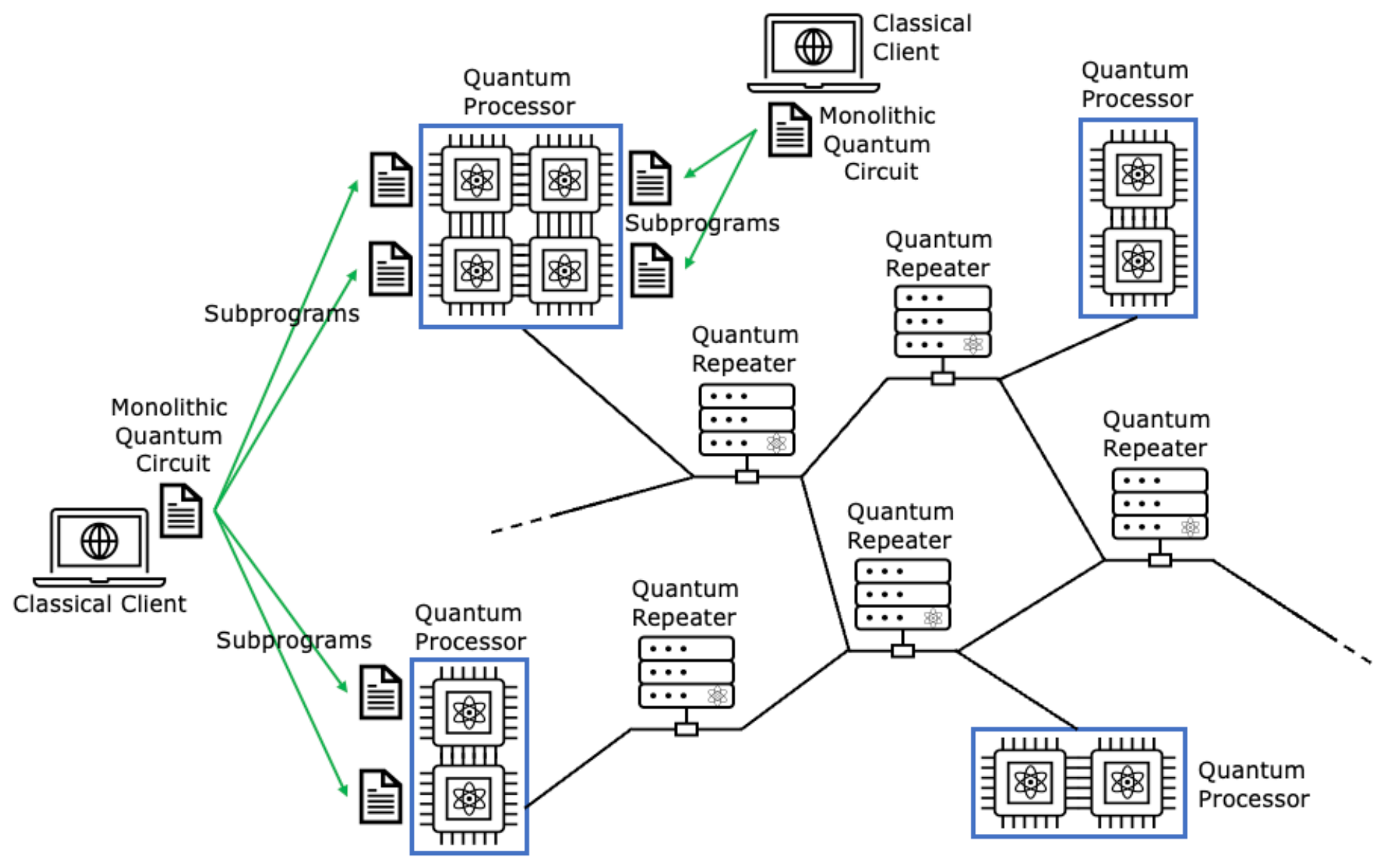}
    \caption{In this example, two jobs are submitted to a geographical-scale DQC system, each one by a different client. The jobs are queued and served in some order. The first job, consisting of two subprograms, is assigned to a quantum processor provided with four QPUs. The second job, consisting of four subprograms, is assigned in part to the aforementioned quantum processor and in part to another quantum processor equipped with two QPUs. The described policy aims at using less quantum processors as possible, occupying as many QPUs as possible.}
    \label{fig:DQC}
    \hrulefill
\end{figure*}

In Fig.~\ref{fig:DQC}, a quantum computation represented by a monolithic circuit gets split into subprograms that are spread for execution over networked QPUs. The subprograms may interact because of \textit{remote gates}, i.e., quantum operators acting on qubits that belong to different nodes.
Quantum compilation issues (i.e, circuit splitting with remote gate minimization and hardware-specific optimization) are out of the scope of this paper. However, the reader must be aware that quantum compilation has a major impact on the quality of the distributed execution~\cite{ZhoChaBie-21,FerCarAmo2023}. 

\begin{figure}[ht!]
    \centering
    \includegraphics[width=\columnwidth]{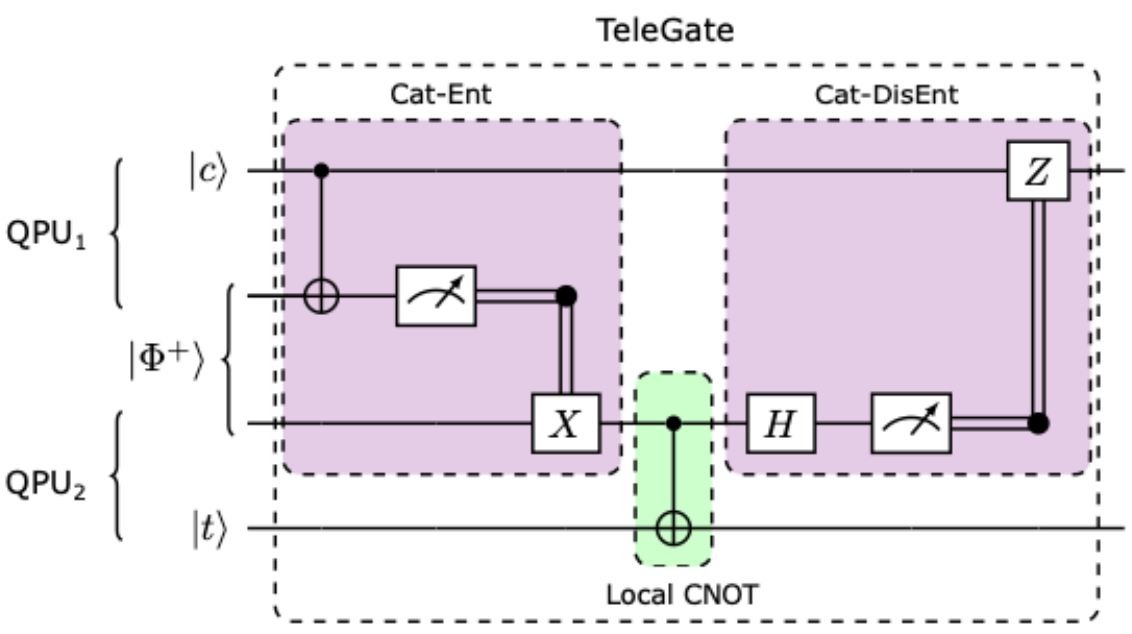}
    \caption{A \texttt{TeleGate} enables a direct gate between remote physical qubits stored at different processors without the need of quantum state teleportation, as long as a Bell state such as $\ket{\Phi^{\texttt{+}}}$ is distributed through the quantum link. For instance, a remote \texttt{CNOT} between $c$ and $t$ can be implemented with two local \texttt{CNOT}s at each processor, followed by conditional gates depending on the measurement of the entangled qubits.}
    \label{fig:tele_gate}
    \hrulefill
\end{figure}

One of the possible ways to perform remote gates is through \texttt{TeleGate}s. As depicted in Fig.~\ref{fig:tele_gate}, a \texttt{TeleGate} enables the execution of a gate between qubits belonging to different QPUs by exploiting a shared entangled pair. For instance, a remote \texttt{CNOT} can be executed across two QPUs at the cost of one entangled pair and two rounds of classical communication. The \texttt{CNOT} is not the only two-qubit operation that can be implemented in this manner, in fact any two-qubit operation can be executed in a remote fashion~\cite{Yimsiriwattana2005}.

Depending on the network scale, DQC poses different challenges. In the case of multi-chip quantum processors, it is highly relevant to prevent bottlenecks when quantum states move across chips. Efficient routing and scheduling algorithms are thus necessary. In current systems, routing operations are set by the compiler and executed by the network layer \cite{Escofet2023}. Such a co-dependency would not make sense with geographically-distributed quantum devices, where it is reasonable to assume that any QPU can interact with any other QPU. I.e., entangled pairs must be established between any two QPUs, when a remote gate must be executed. In this context, entanglement routing is a service provided by the network. Preparing high-quality and high-rate entangled pairs between distant nodes is a major challenge.
Concerning integrated HPC-DQC systems, there are certainly infrastructure challenges. Leveraging the control systems and communication infrastructure already in place could facilitate the installation and management of quantum computers. However, this would necessitate additional components and enhanced control and management measures due to the unique environmental requirements and specialized control systems of quantum machines.
Moreover, developing integrated scheduling systems capable of coordinating the execution of distributed quantum algorithms is imperative. This includes both fully quantum and quantum-classical hybrid algorithms. The latter would particularly benefit from streamlined communication between quantum processors and classical processors.

In this paper, a design framework for DQC simulation is proposed, which puts equal emphasis on computational and networking aspects. Being a design framework, its components can be implemented from scratch, but nothing prevents from reusing existing tools, if they fit the functional requirements. 
The core component of the design framework is an Execution Manager that schedules DQC jobs for running on networked quantum computers. Two metrics are proposed for evaluating the impact of the job scheduling algorithms with respect to QPU utilization and quantum network utilization, beyond the traditional concept of makespan. The discussion is supported by a DQC job scheduling example, where two different strategies are compared in terms of the proposed metrics.

The paper is organized as follows. In Section \ref{sec:motivations}, the motivations of this work are discussed. In Section \ref{sec:framework}, the design framework for DQC simulation is presented. In section \ref{sec:example}, the DQC job scheduling example is illustrated. In section \ref{sec:discussion}, the main open research problems are discussed. Finally, Section \ref{sec:conclusion} concludes the paper with a final recap and some ideas for future work.

\section{Motivations}
\label{sec:motivations}

Many simulation tools have been recently developed to support the research community in the design and evaluation of quantum computing and quantum network technologies, including hardware, protocols and applications \cite{Bartlett2018,Coopmans2021,Dahlberg2018,Matsuo2021,DiAdamoNotzel2021,DahVecDon-22,QNEADK-22}. However, a framework for DQC simulation putting equal emphasis on computational and networking aspects has never been proposed, so far.

Compared to analytical tools, which are well suited to predict the performance of simplified versions of the scenarios of interest, simulation tools find their role in complex use cases. This general concept is particularly true in the quantum computing and networking domain.
Indeed, simulation tools enable the definition of hardware requirements using a top-down approach, that is, starting from applications and protocols. Therefore, high-level performance metrics drive hardware design, which is much more convenient than trial and error. Another advantage of simulation is related to network sizing. Different network topologies and entanglement routing schemes can be devised, given the number of potential users and the number of available quantum processors. This results in saving time and money. 

Regarding DQC, simulation is crucial for establishing the correctness of compiled distributed quantum programs, and evaluating the quality of their execution against different network configurations, hardware platforms and scheduling algorithms. In a recent survey dedicated to DQC, Caleffi et al.~\cite{DQCSurvey2022} compared some prominent simulation tools that were not developed for evaluating DQC systems specifically, but could be used to, with some effort. Each tool is classified as belonging to one of three possible classes: 
\begin{itemize}
    \item hardware-oriented (HW) simulation tools, allowing the user for modeling the physical entities with the desired degree of detail, including noise models;
    \item protocol-oriented (PR) simulation tools, which are mostly devoted to the design and evaluation of general-purpose quantum protocols, -- such as quantum state teleportation, quantum leader election, etc. -- with the possibility to model hardware-agnostic networked quantum processors, with very limited (if not missing) support for noise modeling;
    \item application-oriented (AP) simulation tools, which are tailored to the design and implementation of quantum network applications, relying on simulated backends offered by other packages that are not directly accessible to the user.
\end{itemize}

To the best of our knowledge, the first and -- so far -- only software framework specific for DQC simulation is Interlin-q by Parekh et al.~\cite{Parekh2021}, falling into the AP class. This discrete event simulation platform includes three components, namely the Client Node, the Controller Node and the set of Computing Nodes. The purpose of the Client Node is allowing the user to define a monolithic circuit and the merging function (that is, the function that will be used to merge the partial results of the distributed computation). The Controller Node compiles the monolithic circuit and produces a schedule of quantum programs (that is, subprograms of the monolithic one), taking into account the network topology and the architectures of the quantum processors. Computing Nodes are simulated quantum processors that mimic the execution of the scheduled quantum programs. Such a scheme is effective for the simulation of controlled quantum computing environments, while it lacks the flexibility that is required for simulating DQC over complex quantum networks.

\section{Design Framework for DQC Simulation}
\label{sec:framework}

\begin{figure}[ht!]
    \centering
    \includegraphics[width=.95\linewidth]{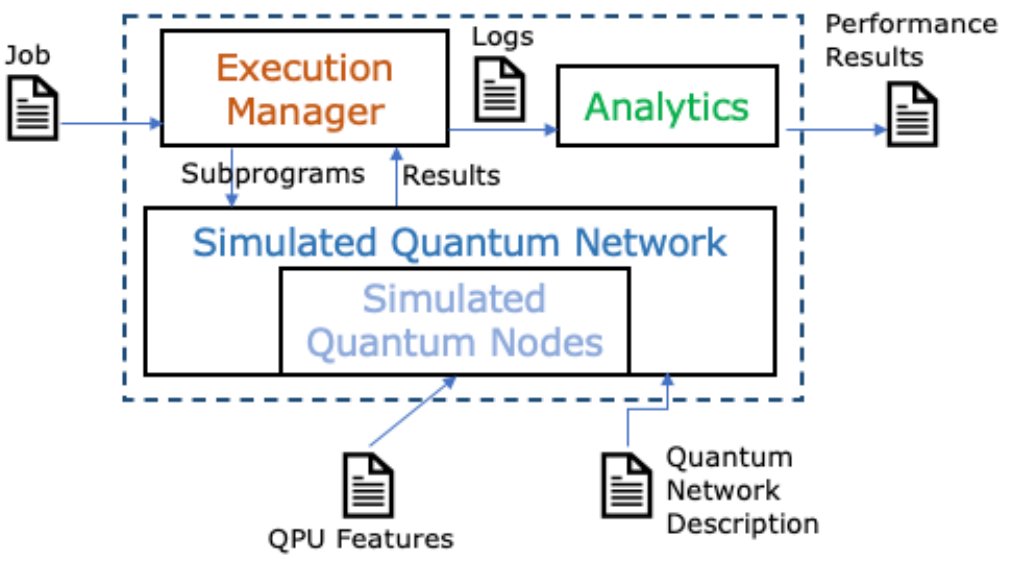}
    \caption{The design framework for DQC simulation.}
    \label{fig:SF}
    \hrulefill
\end{figure}

The proposed design framework includes four components, namely: \textit{Execution Manager},  \textit{Simulated Quantum Network}, \textit{Simulated Quantum Nodes} and \textit{Analytics}. The interconnections between these components are illustrated in Fig. \ref{fig:SF}. The purpose of the Execution Manager is to schedule jobs for concurrent execution on a Simulated Quantum Network, whose Simulated Quantum Nodes mimic the behavior of QPUs, based on low-level descriptions that include detailed qubit and gate models. Finally, the component denoted as Analytics calculates performance metrics from simulation logs collected by the Execution Manager. Further details about these components are provided in the following sections. 

\subsection{Execution Manager}

The Execution Manager is characterized by a queue of jobs to be handled. A job is a quantum circuit compiled for distributed execution. That is, a circuit that has been split into subprograms whose interactions are defined in terms of remote gates that require shared entangled pairs and classical communication between different nodes. 

With respect to classical jobs, DQC jobs have many more static properties that the scheduling policy can exploit. For example, each job is characterized by fixed values of width (i.e., number of data qubits), depth (i.e., number of layers of computation), computation-to-communication ratio (i.e., the ratio between the number of local gates and the number of remote gates), and more. Therefore, the Execution Manager can make clever decisions on the order of execution of the jobs.

More formally, the Execution Manager deals with the \textit{parallel job scheduling} problem, a widely investigated optimization problem in which a set of jobs of varying processing times need to be scheduled on $m$ machines while trying to minimize the makespan, i.e., the length of the schedule. Each job has a processing time $p_i$ (for quantum circuits, this can be approximated with the number of layers of computations) and requires the simultaneous use of $q_i$ machines.
In general, the problem is NP-hard. List-scheduling is an efficient greedy algorithm that guarantees a makespan that is always at most $2-1/m$ multiplied by the optimal makespan \cite{Berit2006}. List-scheduling also works in the online setting where jobs arrive over time and the length of a job becomes known only when it completes \cite{Sgall2023}. 

Fair-share queue policies may be defined as well, to prevent individual users to monopolize resources. For example, jobs may be given different priorities depending on the amount of quantum resources consumed by their owners in the current week. Therefore, users that submit large jobs at a fast pace will face rapidly growing waiting times for their queued jobs, because of decreasing priorities. 

\subsection{Simulated Quantum Network}

In the cases of multi-chip quantum processors and integrated DQC-HPC facilities, direct quantum links can be established between neighbor QPUs. The topology cannot be fully connected, for practical reasons and for the sake of scalability. Therefore, to entangle the qubits of remote QPUs that are not directly connected, it is necessary to use intermediate QPUs as quantum repeaters \cite{FerCacAmo2021}. Moreover, classical communications are characterized by large bandwidth and low latency, making high-performance control messaging an almost obvious assumption.  
As a consequence of these properties, network simulation is not particularly complex in the aforementioned cases.

Instead, in the case of geographically-distributed DQC, according to the vision of the Quantum Internet~\cite{Wehner2018}, quantum links can be created between any two QPUs, leveraging quantum repeaters~\cite{ManAmo2022}. Entanglement generation (including purification) and routing require local quantum operations and classical communication. At the global scale, the impact of classical communication latency cannot be neglected. Therefore, network simulation is more complex, in this case.
In principle, one could simulate the network layers altogether: physical (channels, light-matter interfaces), link (coping with protocols for generating entanglement between neighbor nodes), network (end-to-end entanglement). In practice, what really matters for DQC simulation is the precise characterization of end-to-end entanglement quality and rate. Therefore, separate simulation of the network layers makes more sense, with metrics passed between layers to finally achieve the desired end-to-end performance.

\subsection{Simulated Quantum Nodes}

In principle, a DQC system could integrate heterogeneous QPUs, based on different quantum technologies. Nowadays, the most advanced ones are superconducting qubits (used, for example, by IBM, Google, and Rigetti), trapped ions (used by AQT, Quantinuum and IonQ), neutral atoms (used by PasQal, QuEra, and Atom Computing), and photons (used by Quandela and Xanadu). Other promising quantum technologies are emerging, such as molecular nanomagnets \cite{Chiesa2023} (which are particularly suitable to act as multilevel logical units, i.e., qudits). 

To cope with such a varied scenario, a modular simulator is needed, allowing for the inclusion of detailed physical layer models. A remarkable example is NetSquid \cite{Coopmans2021}, a discrete-event based platform for simulating various aspects of quantum networks, encompassing the physical layer, its control plane, and extending up to the application level. It accomplishes this by integrating several technologies, including a discrete-event simulation engine, a specialized quantum computing library, a modular framework for representing quantum hardware devices, and an asynchronous programming framework to describe quantum protocols. A notable feature of NetSquid is its capacity to accurately take into account the impact of time on the performance of quantum network and quantum computing systems. This is essential in developing scalable systems that can cope with the short lifespan of qubits in quantum devices.

\subsection{Analytics}

Simulation allows evaluating several different aspects of DQC systems. First of all, the correctness of a compiled circuit. Indeed, by simulating its execution on an ideal quantum network, it is possible to analyze the output of the distributed computation, consisting in a quantum state and a classical binary string, respectively before and after the final measurement of the qubits. By repeating the same simulation several times, the obtained output can be compared to the desired one by means of two performance metrics, namely the \textit{classical fidelity} and the \textit{quantum fidelity}. 

Classical fidelity (usually denoted as Hellinger fidelity and related to, although different from, the Bhattacharyya coefficient) is a measure of the amount of overlap between two statistical samples or populations. It thus allow comparing the probability mass function of the classical output to the desired one.

Quantum fidelity is a popular measure of closeness between density operators, which are matrices that describe quantum states more generally than state vectors or wavefunctions. While those can only represent pure states, density matrices can also represent mixed states. In case of multipartite quantum systems (like DQC ones), calculating the partial trace of the global density operator results in a reduced density operator characterizing the quantum state of the subsystem of interest (for example, the quantum state of the qubits in one specific QPU). In this way, if the simulation tool does not allow to evaluate the fidelity of the global quantum state spread across the network (which may be computationally expensive), it is possible at least to evaluate the fidelities of its pieces.

More specific to DQC is the performance evaluation of the Execution Manager. Here two metrics are proposed that go beyond the traditional concept of makespan, namely the \textit{QPU utilization} and the \textit{quantum network utilization}. 

QPU utilization is defined as the ratio between the actual workload and the potential workload given the makespan $M$, namely:
\begin{equation}
U_{\text{QPU}} = \frac{\sum_i p_iq_i} {M n_{\text{QPU}}} \in [0,1], 
\end{equation}

where $p_i$ is the length of the $i$-th job (approximated by the depth of the quantum circuit), $q_i$ is the number of required QPUs of the $i$-th job, $M$ is the makespan of the schedule, and $n_{\text{QPU}}$ is the number of the system's QPUs. For the same job load, different job scheduling algorithms may produce different makespans. The shorter the makespan, the higher the QPU utilization. 

Quantum network utilization is defined as the ratio between the total number of scheduled remote gates and the maximum number of remote gates given the makespan $M$, namely:
\begin{equation}
U_{\text{QN}} = \frac{\sum_i N_{\text{R}i}}{\frac{(n_{\text{QPU}} - 1)M}{r}} =  \frac{r \sum_i N_{\text{R}i}}{(n_{\text{QPU}}-1) M} \in [0,1], 
\end{equation}

where $r$ is the number of layers that are necessary to implement a remote gate (as seen in Fig. \ref{fig:tele_gate}, the \texttt{TeleGate} solution has $r=7$), $N_{\text{R}i}$ is number of remote gates in the $i$-th job, $n_{\text{QPU}}-1$ is the maximum number of remote gates in a layer that spans all the system's QPUs, and $M$ is the makespan of the schedule. Also in this case, the shorter the makespan, the higher the quantum network utilization.

Other relevant DQC aspects that can be evaluated by means of simulation, are the consequences of the non-ideal quantum network over which the distributed quantum computation is performed. Indeed, using realistic models for classical and quantum channels for a given quantum network, it is possible to thoroughly compare different DQC strategies (concerning compilation, job scheduling, output merging) and try to push DQC to the limit. Clearly, there is a trade-off between the rate of executed computations and the quality of their outputs. On the other hand, simulated DQC is also a great opportunity for quantum network designer to learn what hardware components and protocols need to be improved first.

In the next section, the proposed utilization metrics are further analyzed by means of a simple example.

\section{DQC Job Scheduling Example}
\label{sec:example}

In this section, a DQC scenario is presented that can be simulated following the design framework described in Section 3. The Execution Manager implements both first-in first-out (FIFO) and list-scheduling algorithms, for scheduling the execution of distributed quantum computations. DQC at geographical scale is assumed, so that entanglement between any two QPUs is provided by the network.

Algorithm 1 is the adopted list-scheduling strategy, which is repeated every time a job execution completes.

\begin{algorithm}[ht!]
    \caption{List-Scheduling
        \newline
        \footnotesize
        \textbf{Input}: job queue $J$, idle QPU set $Q$
        \newline
        \textbf{Output}: 
    }
    \label{alg:schedule}
    \begin{algorithmic}[1]
        \Function{Schedule}{}
        \State $i \gets 0$
        \While{$Q \neq \emptyset$}
            \State $next \gets J[i]$
            \If{$\exists q \subseteq Q : q = next.q$}
                \State schedule $next$
                \State $Q \gets Q \char`\\ q$
                \State $J \gets J \char`\\ next$
            \Else
            \State $i \gets i+1$
            \EndIf
        \EndWhile
        \EndFunction
    \end{algorithmic}
\end{algorithm}

\begin{figure*}[ht!]
    \centering
    \includegraphics[width=\textwidth]{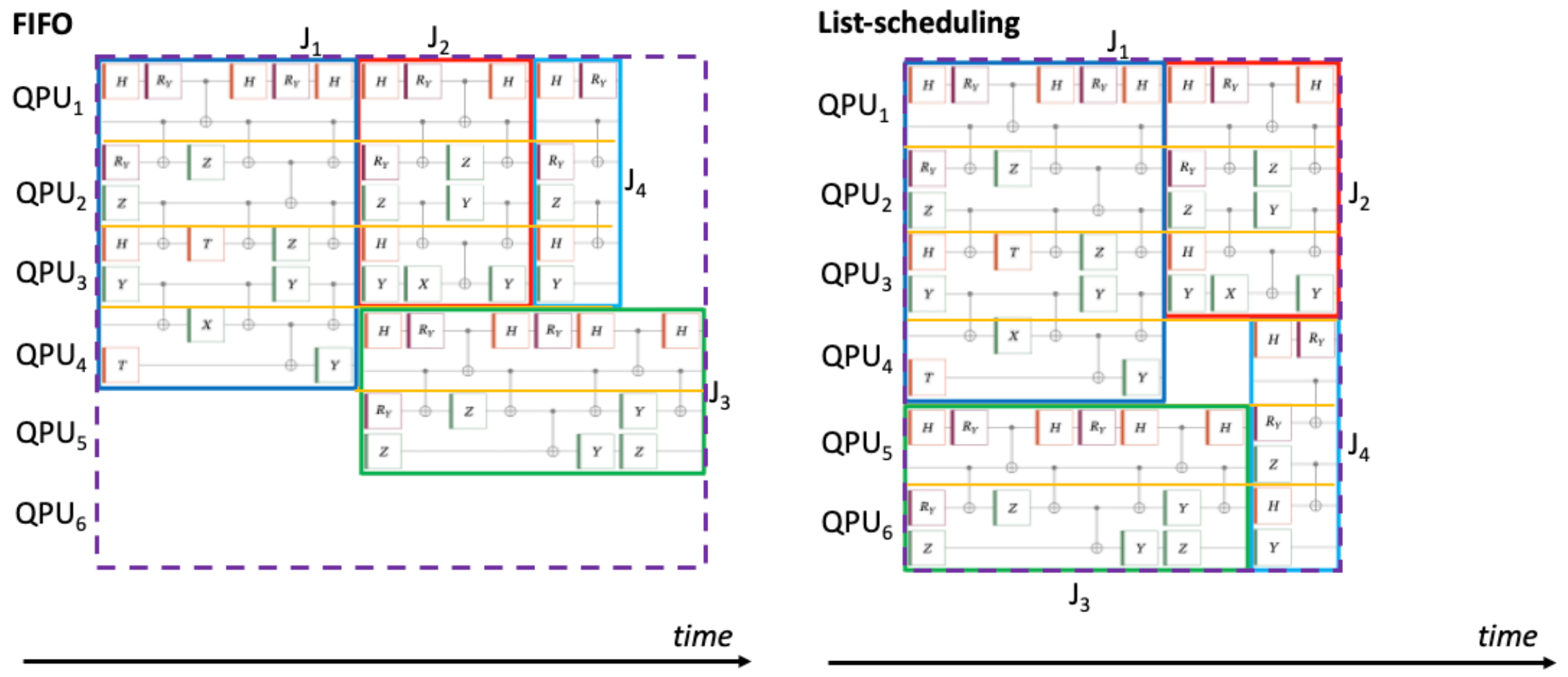}
    \caption{Schedules produced by the FIFO and list-scheduling algorithms, considering the four jobs illustrated in Table \ref{tab:features} and a DQC system with $n_{\text{QPU}} = 6$. The dashed line surrounds an area whose value is $M n_{\text{QPU}}$, where $M$ is the makespan of the schedule. Such a product is the denominator of the $U_{\text{QPU}}$ formula. The area occupied by the scheduled jobs is the numerator, instead. Therefore, the difference in QPU utilization between the two schedules has also a visual evidence.}
    \label{fig:schedules}
    \hrulefill
\end{figure*}

In this example, there are four jobs $\{\text{J}_1,\text{J}_2,\text{J}_3,\text{J}_4\}$ that must be scheduled on a six-QPU system ($n_{\text{QPU}} = 6$). Each job $\text{J}_i$ is characterized by an arrival number $i$ ($i=1$ meaning first arrived), a length $p$, a number of requested QPUs $q$, a number of remote gates $N_\text{R}$, and a number of local gates $N_\text{L}$. A derived job feature is the Computation-to-Communication Ratio (CCR), defined as the ratio between $N_\text{L}$ and $N_\text{R}$. The features of the four jobs are reported in Table \ref{tab:features}.

\begin{table}[h!]
\begin{center}
\caption{Features of the four jobs $\{\text{J}_1,\text{J}_2,\text{J}_3,\text{J}_4\}$ that must be scheduled on a six-QPU system, in the proposed example.}
\begin{tabular}{ |c|c|c|c|c|c|c| } 
\hline
Job Name & $i$ & $p$ & $q$ & $N_R$ & $N_L$ & $CCR$\\
\hline
\hline
J$_1$ & 1 & 24 & 4 & 9 & 19 & 2.11\\ 
\hline
J$_2$ & 2 & 16 & 3 & 4 & 13 & 3.25\\ 
\hline
J$_3$ & 3 & 32 & 2 & 4 & 15 & 3.75\\ 
\hline
J$_4$ & 4 & 8 & 3 & 2 & 6 & 3\\ 
\hline
\end{tabular}
\label{tab:features}
\end{center}
\end{table}

Fig. \ref{fig:schedules} compares the schedules produced by the FIFO and list-scheduling algorithms. For the ease of reading, remote gates are represented as single gates. However, as previously shown in Fig. \ref{fig:tele_gate}, the compiler turns each 1-layer remote gate into a 7-layer circuit. Therefore, the actual depths of the circuits illustrated in Fig. \ref{fig:schedules} must be calculated as the sum of the number of layers with local gates only $p_\text{L}$ and the number of layers with local and remote gates $p_\text{R}$.

Using the FIFO algorithm, the makespan is $M_{\text{FIFO}} = 56$ and the QPU utilization is $U^{\text{FIFO}}_{\text{QPU}} = 0.69$.
With the list-scheduling algorithm, the makespan is $M_{\text{LS}} = 40$ and the QPU utilization is $U^{\text{LS}}_{\text{QPU}} = 0.96$. Actually, the list-scheduling algorithm produces a schedule that is optimal in terms of makespan, with all six QPUs being almost fully utilized.

In a classical computing setting, optimal makespan and full resource utilization would be highly appreciated. In distributed quantum computing, the story is quite different. Indeed, makespan optimality needs highly effective and efficient entanglement routing between QPUs, in order to guarantee timely execution of remote gates that are all concentrated in a short time frame. Indeed, with the FIFO algorithm, the network utilization is $U^{\text{FIFO}}_{\text{QN}} = 0.475$, while using the list-scheduling algorithm, the network utilization is $U^{\text{LS}}_{\text{QN}} = 0.665$. Once more, it is worth remembering that the number of layers that are necessary to implement a remote gate (denoted as $r$) is assumed to be equal to $7$. 

\vspace{-0.1cm}
\section{Discussion}
\label{sec:discussion}

With current technologies, searching for a tradeoff between QPU and quantum network utilization is crucial. The practical implementation on real quantum hardware imposes several specific constraints, hindering the effectiveness of simplistic job scheduling algorithms. Furthermore, if multiple subprograms are scheduled simultaneously on the same QPU (quantum multiprogramming), crosstalk will be a relevant issue. Recent works \cite{Ohkura2022, Khadir2023, Niu2023} proposed greedy mappers that seek to minimize crosstalk and number of gates by leveraging ``buffers'' (i.e., unused qubits between allocated quantum circuits) and iteratively trying different mappings. These approaches trade crosstalk avoidance with resource utilization and take the risk of getting stuck into local minima.

In summary, finding effective, efficient and flexible DQC-specific job scheduling algorithms is an interesting open research problem. Which is even more challenging when considering heterogeneous DQC systems, with QPUs and network devices based on different quantum technologies.

\section{Conclusion and Future Work}
\label{sec:conclusion}
In this work, a design framework for DQC simulation was proposed, together with two metrics for evaluating job scheduling algorithms in terms of QPU utilization and quantum network utilization. The discussion was supported by a DQC job scheduling example, where two different strategies are compared in terms of the proposed metrics. The need for novel DQC-specific job scheduling algorithms emerged as an important open research problem.

There is a quite variegated choice of simulation tools for studying either quantum networks or quantum computers that may be used to support DQC research. 
Based on the proposed design framework, future work will concern the study and development of modules that allow for bridging existing simulation tools, with automated instantiation of simulation objects representing QPUs and quantum network components.

\section*{Acknowledgement}
Davide Ferrari acknowledges financial support from the EU Flagship on Quantum Technologies through the project Quantum Internet Alliance (EU Horizon Europe, grant agreement no. 101102140).
Michele Amoretti acknowledges financial support from the European Union – NextGenerationEU, PNRR MUR project PE0000023-NQSTI.

\bibliographystyle{IEEEtran}
\bibliography{bibliography.bib}

\end{document}